\documentclass[aps,prl,twocolumn,showpacs,showkeys,superscriptaddress,longbibliography,nobalancelastpage]{revtex4-1}

\usepackage{graphicx}
\usepackage{amsmath}
\usepackage{amssymb}
\usepackage{bm}
\usepackage{hyperref}
\usepackage{color}
\hypersetup{colorlinks=true,citecolor=black,linkcolor=black,urlcolor=black}

\begin{document}

\title{Spin fluctuations associated with the collapse of the pseudogap in a cuprate superconductor}

\author{M. Zhu}
\affiliation{H.H. Wills Physics Laboratory, University of Bristol, Tyndall Avenue, Bristol BS8 1TL, United Kingdom}

\author{D. J. Voneshen}
\affiliation{ISIS Facility, Rutherford Appleton Laboratory, Didcot OX11 0QX, United Kingdom}
\affiliation{Department of Physics, Royal Holloway University of London, Egham, TW20 0EX, United Kingdom}
\author{S. Raymond}
\affiliation{Univ. Grenoble Alpes, CEA, IRIG, MEM, MDN, 38000 Grenoble, France}
\author{O. J. Lipscombe}
\affiliation{H.H. Wills Physics Laboratory, University of Bristol, Tyndall Avenue, Bristol BS8 1TL, United Kingdom}
\author{C. C. Tam}
\affiliation{H.H. Wills Physics Laboratory, University of Bristol, Tyndall Avenue, Bristol BS8 1TL, United Kingdom}
 \affiliation{Diamond Light Source, Harwell Campus, Didcot OX11 0DE, United Kingdom.}
\author{S. M. Hayden}
\email{s.hayden@bris.ac.uk}
\affiliation{H.H. Wills Physics Laboratory, University of Bristol, Tyndall Avenue, Bristol BS8 1TL, United Kingdom}

\begin{abstract}
Theories of the origin of superconductivity in cuprates are dependent on an understanding of their normal state which exhibits various competing orders.  
Transport and thermodynamic measurements on La$_{2-x}$Sr$_x$CuO$_4$ show signatures of a quantum critical point, including a peak in the electronic specific heat $C$ versus doping $p$, near the doping $p^{\star}$ where the pseudogap collapses.  The fundamental nature of the fluctuations associated with this peak is unclear. Here we use inelastic neutron scattering to show that close to $T_c$ and near $p^{\star}$, there are very-low-energy collective spin excitations with characteristic energies $\hbar \Gamma \approx$~5 meV. Cooling and applying a 8.8~T magnetic field creates a mixed state with a stronger magnetic response below 10~meV.  We conclude that the low-energy spin-fluctuations are due to the collapse of the pseudogap combined with an underlying tendency to magnetic order. We show that the large specific heat near $p^{\star}$ can be understood in terms of collective spin fluctuations. The spin fluctuations we measure exist across the superconducting phase diagram and may be related to the strange metal behaviour observed in overdoped cuprates.

\end{abstract}

%\pacs{74.72.$−$h,25.40.Fq}
%\keywords{}

\maketitle

%------------------------------------------------------------------------------------------
%\section
% Spin fluctuations in cuprates

Spin fluctuations can play an important part in determining the low-temperature thermal and quasiparticle properties of strongly-correlated electron systems.  Notable examples are heavy-fermion metals \cite{Stewart1984_S,Coleman2015_C} such as CeCu$_6$ and UPt$_3$.  At low temperatures, these materials show very-large linear heat capacities $\gamma=C/T$ because they form heavy electron quasiparticles incorporating moments of the $4f$ or $5f$ electrons and low-energy ($\lesssim 1$\;meV) spin fluctuations develop \cite{Walter1986_WWF}. While cuprate superconductors (SC) do not show the very-large quasiparticle mass $m^{\star}$ observed in heavy-fermions systems, they do show moderate enhancements  of $\gamma$ and $m^{\star}$ up to a factor of $\sim$3 with respect to the local-density-approximation (LDA) band structure calculations \cite{Rourke2010_RBBM,Ramshaw2015_RSMD,Yoshida2007_YZLK, Horio2018_HHSM,Markiewicz2005_MSLL}. In this work, we investigate how the spin degrees of freedom contribute to the relatively large $\gamma$ observed for certain dopings in the cuprates.

It is well known that the high-energy spin excitations persist across the superconducting phase diagram of cuprate superconductors \cite{Keimer2015_KKNU}. These excitations can have energies comparable with the exchange constant $J \approx 120$\;meV of the parent antiferromagnets. They are strong near the antiferromagnetic zone centre and are believed to cause superconductive pairing \cite{Scalapino2012_S}. The normal state of cuprate superconductors shows unusual behaviour in transport and thermodynamic properties \cite{Timusk1999aa,Proust2019_PT,Keimer2015_KKNU,Cooper2009aa,Proust2019_PT,Michon2019fk} such as the ``Planckian'' linear $T$-dependence of the resistivity \cite{Keimer2015_KKNU,Cooper2009aa,Hartnoll2021_HM}. These properties are related to excitations with lower energies comparable with $k_B T$ rather than $J$.

\begin{figure}[htb]
\includegraphics[width = 0.45\textwidth]{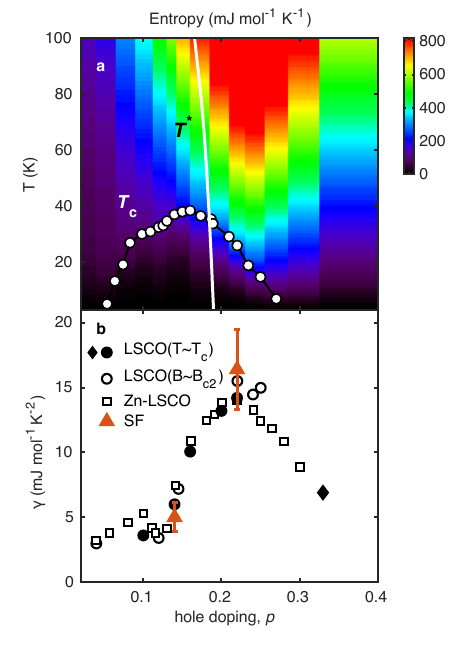}
\caption{\textbf{Entropy and electronic specific heat in La$_{2-x}$Sr$_x$CuO$_4$. } \textbf{a} Entropy as a function of temperature and hole doping $p=x$ for LSCO (derived from data of Ref.~\cite{Loram2001xp}). $T_c(p)$ is the superconducting critical temperature (open circles) \cite{Takagi1989}. $T^{\ast}(p)$ (solid line) is the pseudogap temperature \cite{Timusk1999aa}.  
\textbf{b} Doping dependence of the electronic specific heat coefficient $\gamma(T) = C_{\text{el}}/T$ in the normal state, for $T \approx T_c$ (closed circles \cite{Matsuzaki2004_MMOI}, closed diamond \cite{Nakamae2003_NBMN}), at high magnetic field $B \approx B_{c2}$ (open circles \cite{Girod2021_GLDS}), or where superconductivity is suppressed by Zn doping (open squares \cite{Momono1994_MINO}).   
The solid triangles represent $\gamma(T=T_c)$ of LSCO ($p$ = 0.14 and 0.22) calculated using the spin fluctuation theory described in the text.
\label{fig:entropy_heat_capacity}}
\end{figure}

The single-layer cuprate superconductor La$_{2-x}$Sr$_x$CuO$_4$ (LSCO) can be doped across the superconducting phase diagram, where the hole doping $p=x$.  Normal-state heat-capacity measurements have been made at $T \approx T_c$  \cite{Loram2001xp, Matsuzaki2004_MMOI} and also at lower temperatures with superconductivity suppressed by Zn doping  \cite{Momono1994_MINO}, a high magnetic field \cite{Girod2021_GLDS}, or a high magnetic field and Nd/Eu doping \cite{Michon2019fk}. It is found that the specific heat $\gamma(p)$ shows a peak at $p=p_c \approx 0.22$ (see Fig.~\ref{fig:entropy_heat_capacity}b).   The peak in $\gamma(p)$ and the fan-shaped entropy landscape (Fig.~\ref{fig:entropy_heat_capacity}a) above $T_c$ resemble systems such as iron-based superconductors \cite{Shibauchi2014xq} and Sr$_3$Ru$_2$O$_7$ \cite{Lester2021_LRPC} that display magnetic quantum criticality and where enhancements in the quasiparticle mass have been associated with the presence of low-energy spin fluctuations.  Thus, it is natural to ask whether spin fluctuations contribute to the large $\gamma$ values near $p_c$ in LSCO. The situation in LSCO is subtle because the Fermi energy $E_F$ passes through a van-Hove singularity (VHS) \cite{Yoshida2007_YZLK,Horio2018_HHSM} near $p_c$ as the doping is increased and the pseudogap (PG) \cite{Timusk1999aa} terminates at a critical doping  $p^{\star} \approx 0.19$ \cite{Cooper2009aa,Girod2021_GLDS} which is close to $p_c$.    

\begin{figure*}[htb]
\includegraphics[width = \textwidth]{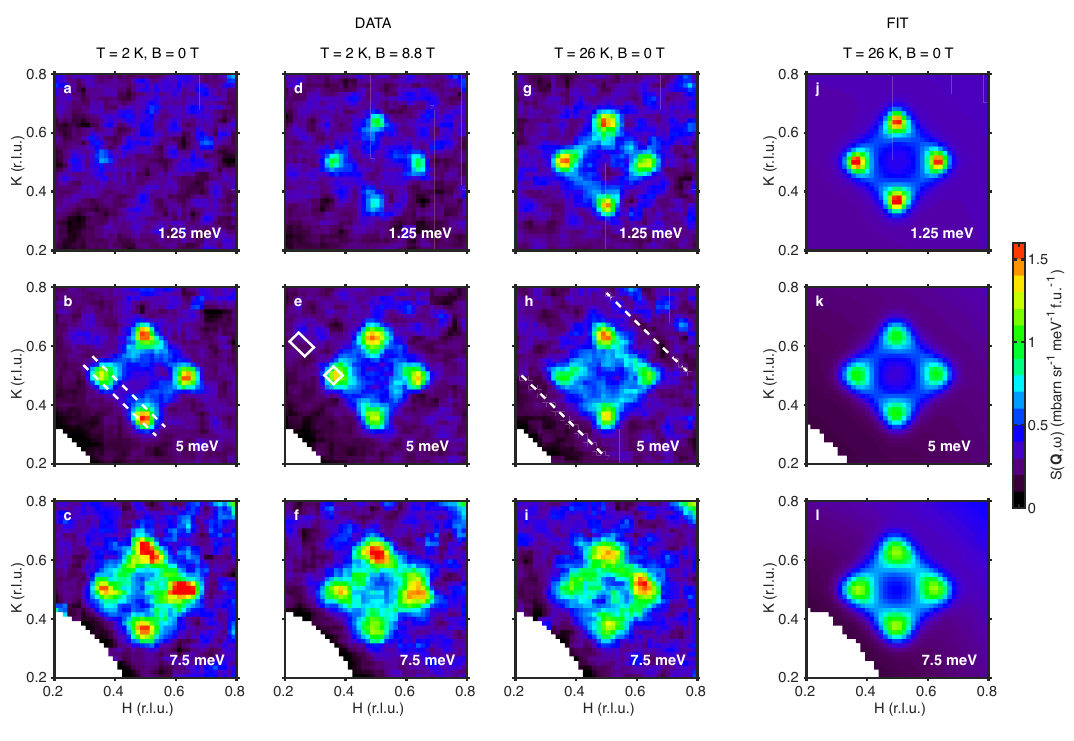}
\caption{\textbf{Wavevector-dependent maps of low-energy spin fluctuations in La$_{2-x}$Sr$_x$CuO$_4$ ($x = 0.22$).} Constant-energy maps of $S(\mathbf{Q},\omega)$ measured at: \textbf{a-c} $T$ = 2 K, $B$ = 0 T; \textbf{d-f} $T$ = 2 K, $B$ = 8.8 T; \textbf{g-i} $T$ = 26 K ($T_c$), $B$ = 0 T. $L$ is integrated over the range $|L| \le 1$. White dashed lines in \textbf{b} are the range of integration used to produce Fig.\;\ref{fig:LET_q_cuts}. White boxes in \textbf{e} define integration ranges for signal and background used to produce Fig.\;\ref{fig:LET_E_cuts}a-b. 
\textbf{j-l} The result of a global fit of the Eqn.\;\ref{eqn:chi_MMP_SM} including correction of a magnetic form factor and a $|\mathbf{Q}|^2$ background. Data shown in Figs.~\ref{fig:LET_slices}-\ref{fig:LET_E_cuts} were collected on LET.
\label{fig:LET_slices}}
\end{figure*}

%In the cuprates with high $T_c \approx 90$~K and large superconducting gaps, such as YBa$_2$Cu$_3$O$_{6+x}$, the spin fluctuations are strongly suppressed \cite{Headings2011_HHKB} in superconducting state and for $\hbar \omega \lesssim 30$\;meV. Thus, the signature of low-energy spin fluctuations is hard to observe. Lower $T_c$ systems such as LSCO offer the opportunity to measure at lower temperatures where low-energy spin fluctuations are well developed.

We have measured the low-energy spin fluctuations in the normal, mixed and superconducting states of LSCO for $p=0.22 \approx p_c$. In the normal state, at $T=T_c=26$\;K, we find incommensurate spin fluctuations with a low energy scale $\hbar \Gamma=4.6 \pm 0.3$\;meV and a correlation length $\xi=19 \pm 2$~\AA. On repeating the measurement in the mixed state created by applying a magnetic field $B=8.8$\;T at $T=2$\;K (i.e. $T \ll T_c$),  we find that the low-energy spin excitations are enhanced for all energies below 10~meV compared with the normal state at $T \sim T_c$.  This indicates that, if superconductivity had not intervened, the low-energy spin fluctuations in the normal state would be substantially stronger at $T$=2~K than at $T \sim T_c$.  Thus the superconducting dome in LSCO hides a region of coherent very low-energy spin fluctuations near $p_c$.  

\begin{figure*}[tb]
\includegraphics[width = \textwidth]{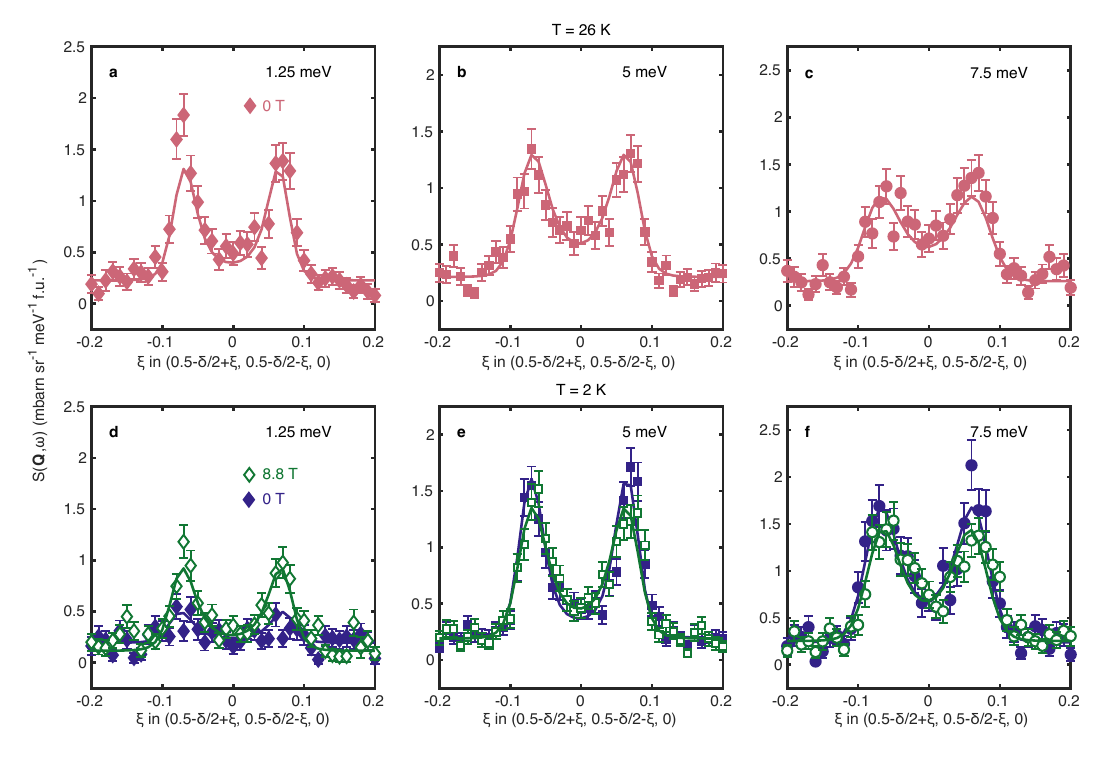}
\caption{\textbf{Fits of spin-fluctuation model to magnetic excitations.} \textbf{a-c} Wavevector-dependent $S(\mathbf{Q},\omega)$ cuts through $\mathbf{Q}_{\delta}$ for various energies in the normal state. Solid lines are the result of a global fit to the phenomenological spin fluctuation model Eqn.~\ref{eqn:chi_MMP_SM}. \textbf{d-f} Cuts in the superconducting state. Low-energy fluctuations are induced by a magnetic field (panel \textbf{d}). Lines are fits to Eqn.~\ref{eqn:4} with $\kappa_1(\omega)=0$ (see Methods). Error bars are determined from Poisson counting statistics or least squares fitting of data and denote one standard deviation.
\label{fig:LET_q_cuts}}
\end{figure*}

%\section{Results} 
\label{sec:3} 
Inelastic neutron scattering (INS) studies \cite{Thurston1989_TBKP,Mason1993_MAHR, Aeppli1997jl, Mook1998_MDHA, Hinkov2004_HPBS, Wakimoto2004aa,Lipscombe2007_LHVM, Li2018ab, Ikeuchi2018aa} 
show that, except for very low dopings, the low-energy, $\hbar \omega \lesssim 30$\;meV, spin excitations in cuprate superconductors are strongest at incommensurate wavevectors $\mathbf{Q_{\delta}}$ = (0.5$\pm\delta$, 0.5) and (0.5, 0.5$\pm\delta$) and occur throughout the phase diagram of materials such as YBa$_2$Cu$_3$O$_{6+x}$ and LSCO.  For both materials, $\delta$ increases with doping \cite{Yamada1998aa}, saturating at $\delta \approx 0.134$ for $p \gtrsim 0.20$ in LSCO. In the superconducting state, these excitations are suppressed at low energies \cite{Mason1993_MAHR,Headings2011_HHKB} approximately below the superconducting gap $\Delta$.  In the normal state of underdoped cuprates, for example LSCO ($p$ = 0.14),  the characteristic energy $\hbar\Gamma$ of excitations is strongly temperature-dependent and $\omega/T$ scaling has been observed \cite{Aeppli1997jl}.  Previous studies \cite{Lee2003_LYHV,Wakimoto2004aa,Lipscombe2007_LHVM,Li2018ab,Ikeuchi2018aa} of overdoped LSCO have identified low-energy incommensurate magnetic scattering. However, a quantitative characterisation of the magnetic response for $p \approx p_c$ in the normal state has not been attempted.  

In this study we use (see Methods) the LET, MERLIN and IN12 spectrometers to map out the \textbf{Q}-$\omega$ dependence of the low-energy spin excitations in LSCO ($p$ = 0.22).  A complication in this measurement is the presence of phonon scattering. LSCO undergoes a tetragonal (HTT) to orthorombic (LTO) structural phase transition for $p \lesssim 0.21$. The soft phonon \cite{Birgeneau1987_BCGJ} associated with this transition has a reduced wavevector $(0.5,0.5,0)$ and an energy of $\sim3$~meV in LSCO ($p$ = 0.22).  The intensity of phonon scattering is proportional to $|\mathbf{Q}|^2$ times a structure factor and in this measurement we minimise the phonon scattering by measuring at small $|\mathbf{Q}|$ near $\mathbf{Q}=(0.5,0.5,L)$ with $|L| \leq 1$ (See Extended Data Figs.~\ref{fig:LET_MERLIN}-\ref{fig:MERLIN_phonons} for further details).  Our samples showed no evidence of incommensurate magnetic order at $T=1.5$~K and $B=10$~T (See Extended Data Fig.~\ref{fig:IN12_noSDW}) in agreement with recent nuclear magnetic resonance measurements \cite{Frachet2020_FVZB} that this only exists for $p<p^{\star}$. We find that any ordered moment would be less than $0.006 \, \mu_B$~Cu$^{-1}$. 

\begin{figure*}[tb]
\includegraphics[width = \textwidth]{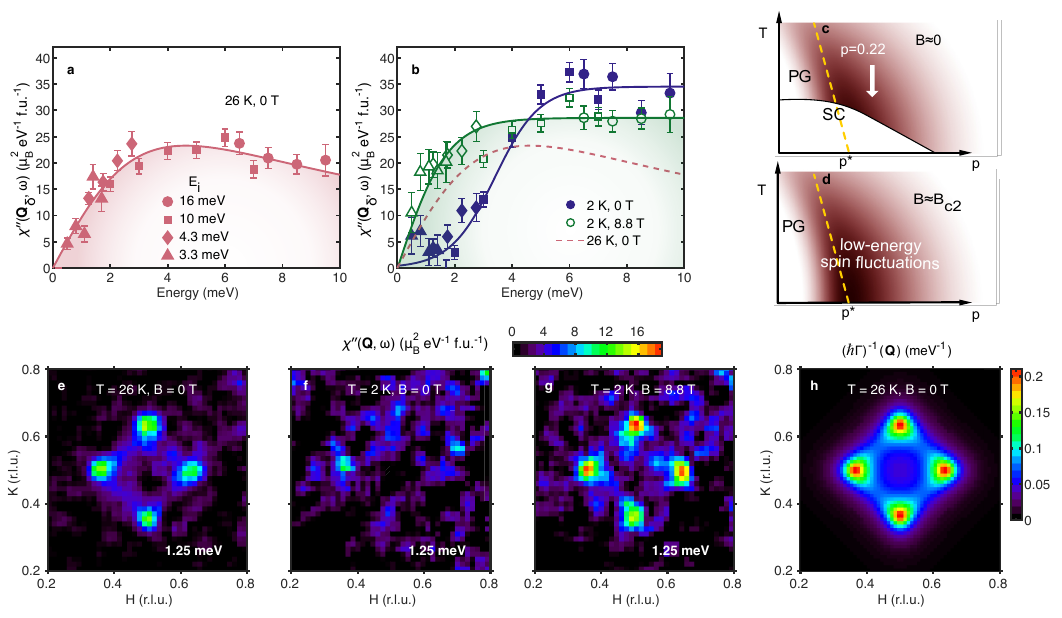}
\caption{\textbf{Magnetic excitations with a low energy scale in the normal and mixed states of LSCO ($p$ = 0.22)}.  Energy dependence of the magnetic response $\chi^{\prime\prime}(\mathbf{Q}_{\delta},\omega)$. \textbf{a} Normal state ($T=26$~K, $B=0$, pink) showing the low energy scale. \textbf{b} Increased response in the mixed state ($T=2$~K, $B=8.8$~T, green, open symbols) and suppression at low energy in the superconducting state ($T=2$~K, $B=0$, blue, closed symbols). The solid line in \textbf{a} is a fit of $\chi^{\prime\prime}(\mathbf{Q}_{\delta},\omega)/\omega$ to a Lorentzian response. Sold lines in \textbf{b} are guides to the eye. Symbols indicate incident energy $E_i$. \textbf{c} Schematic illustrating the magnitude of  $\chi^{\prime\prime}(\mathbf{Q}_{\delta},\hbar\omega \sim \textrm{1 meV})$ (red shading) in the normal state and \textbf{d} its enhancement under applied magnetic field. \textbf{e-g} Maps of low-energy $\chi^{\prime\prime}(\mathbf{Q},\omega)$ at $T \approx T_c$ and at $T=2$~K with $B=0,8.8$~T. \textbf{h} Wavevector-dependence of the relaxation parameter $\Gamma^{-1}(\mathbf{Q})$ in the normal state obtained from a global fit.  
\label{fig:LET_E_cuts}}
\end{figure*}

Figure \ref{fig:LET_slices}a-i show the constant-energy maps of the scattering intensity $S(\mathbf{Q},\omega)$ at $\hbar\omega$ = 1.25, 5 and 7.5 meV measured at $T$ = 26\;K ($T_c$) and 2 K at zero field, and $T$ = 2 K in a magnetic field of $B$ = 8.8 T applied parallel to the $c$ axis.  The magnetic response in this energy range is peaked at the four incommensurate wavevectors $\mathbf{Q}_{\delta}$ with $\delta \approx$ 0.135 in agreement with a previous study at higher energies \cite{Lipscombe2007_LHVM}.
Fig.\;\ref{fig:LET_q_cuts} shows wavevector-dependent $\mathbf{Q}$ cuts of $S(\mathbf{Q},\omega)$ through the $\mathbf{Q}_{\delta}$ positions along the trajectory shown in Fig.~\ref{fig:LET_slices}b. Fig.\;\ref{fig:LET_E_cuts}a,b and Fig.\;\ref{fig:LET_E_cuts}e-g show the imaginary part of the dynamical spin susceptibility $\chi^{\prime\prime}(\mathbf{Q},\omega)$ extracted directly from the measured $S(\mathbf{Q},\omega)$ by correcting for the bose factor (see Methods) and displayed as $\hbar \omega$-cuts at $\mathbf{Q} = \mathbf{Q_{\delta}}$ and $\mathbf{Q}$-slices at $\hbar\omega=1.25$\;meV.  

In Fig.\;\ref{fig:LET_E_cuts}a (pink shading, closed symbols) we see the presence of low-energy spin fluctuations in overdoped LSCO near $p_c$ in the normal state at $T=T_c$. On fitting this $\omega$-dependent cut to a Lorentzian response $\chi^{\prime\prime}(\mathbf{Q}_{\delta},\omega)/\omega \propto \Gamma_{\delta}/(\Gamma^2_{\delta} +\omega^2)$ we find a characteristic energy scale $\hbar\Gamma_{\delta} = 4.6 \pm 0.3$\;meV. To put this in context, we note that for underdoped LSCO ($p=0.14$), where $\omega/T$ scaling is observed \cite{Aeppli1997jl},  a larger $\hbar\Gamma_{\delta} \approx 9.6$\;meV is found at $T=35$\;K $\approx T_c$. Lowering the temperature to 2\;K and applying a modest magnetic field of $B=8.8$~T (Fig.\;\ref{fig:LET_E_cuts}b, green shading, open symbols) creates the mixed state with low-energy excitations associated with the vortices \cite{Lake2001_LACM}. The response is reminiscent of the normal state (Fig.\;\ref{fig:LET_E_cuts}a) with $T \sim T_c$ but $\chi^{\prime\prime}(\mathbf{Q}_{\delta},\omega)$ is enhanced at all energies below 10~meV.  Comparing the superconducting state ($B=0, T=2$~K, Fig.\;\ref{fig:LET_E_cuts}b, blue, closed symbols) with the normal state ($T=26$~K, pink, dashed lines), we see a suppression of the excitations below $\approx 4$\;meV due to the opening of the superconducting gap $\Delta$ and an increase above this energy due to the well-known spin resonance (e.g. Refs.~\cite{Lee2003_LYHV,Headings2011_HHKB}). 

We can parameterize our normal-state data at $T = T_c$ with a phenomenological susceptibility similar to the forms used by Pines \textit{et al.} \cite{Pines1990aa} and Aeppli \textit{et al.} \cite{Aeppli1997jl} 
\begin{equation}
    \chi^{\prime\prime}(\mathbf{Q},\omega) = \frac{\chi_{\delta}\Gamma_{\delta}\omega}{\Gamma_{\delta}^2\left[1+\xi^2 R(\mathbf{Q})\right]^2+\omega^2},
\label{eqn:chi_MMP_SM}
\end{equation}
where $\chi_{\delta}$, $\Gamma_{\delta}$, $\xi$ are independent of $\omega$ and $\mathbf{Q}$, $\chi_{\delta}=\chi^{\prime}(\mathbf{Q}_{\delta},\omega=0)$. $R(\mathbf{Q})$ is a function which has the symmetry of the 2D Brillouin zone and reproduces the four incommensurate peaks, 
\begin{align}
    R(\mathbf{Q}) & = \frac{1}{4 \delta^2} \big\{ [(H-\tfrac{1}{2})^2+(K-\tfrac{1}{2})^2-\delta^2]^2 \big. \nonumber \\
    &\big.  \quad +  4 (H-\tfrac{1}{2})^2(K-\tfrac{1}{2})^2 \big\}.
    \label{eqn:R_Q}
\end{align}
Near $\mathbf{Q}_{\delta}$, $R(\mathbf{Q}) \propto |\mathbf{Q}-\mathbf{Q}_{\delta}|^2$, allowing $\xi$ to be interpreted as a correlation length \cite{Pines1990aa,Aeppli1997jl}.  
Note $\chi^{\prime\prime}(\mathbf{Q},\omega)/\omega \propto 1/[\Gamma^2(\mathbf{Q})+\omega^2]$, this corresponds to an extreme overdamped oscillator (Lorentzian) lineshape (See Supplementary Material and Ref.~\cite{Chaikin1995} for discussion) with a $\mathbf{Q}$-dependent relaxation rate $\Gamma(\mathbf{Q})$  where 
\begin{equation}
    \Gamma(\mathbf{Q}) = \Gamma_{\delta}[1+\xi^2R(\mathbf{Q})].
    \label{eqn:R_Qa}
\end{equation}

Eqn.\;\ref{eqn:chi_MMP_SM} provides a good global (in $\mathbf{Q}$ and $\omega$) description of the excitations in the normal state (see Fig.~\ref{fig:LET_slices}g-l, and Methods). The resulting fitted parameters for $T=26$\;K are shown in Table\;\ref{tab:parms}. For $T=2$\;K in the superconducting state (Fig.\;\ref{fig:LET_E_cuts}b), the line shape is no longer described by the Lorentzian form, however, $\mathbf{Q}$ cuts at a given energy (Fig.~\ref{fig:LET_q_cuts}d-f) can be fitted using the related Eqn.~\ref{eqn:4} (see Methods). 

%Our experiment shows that low-frequency spin excitations with characteristic energy $\hbar \Gamma=4.6 \pm 0.3$\;meV and relatively long correlation lengths $\xi=19 \pm 2$\;\AA\ are present in the normal state ($T=25\;\textrm{K} \approx T_c$) of overdoped LSCO($x=0.22$). 

%\section{Discussion} 
\label{sec:4}
Landau Fermi liquid theory is usually used to understand the low-temperature heat capacity of metals. Quasiparticle states can be renormalised by the interaction with excitations such as spin fluctuations.  In underdoped cuprates the situation is complicated by the pseudogap which removes low-energy quasiparticle states leading to the suppression of $\gamma$ for $p<p^{\star}$  in Fig.~\ref{fig:entropy_heat_capacity}b.

For metals with strong antiferromagnetic or ferromagnetic correlations the low temperature specific heat can also be understood in terms of the spin excitations. The contribution of the spin fluctuations to the free energy can be estimated using a self-consistent renormalisation spin fluctuation (SF) theory based on a one-loop approximation and the Hubbard model \cite{Beal-Monod1968na,Brinkman1968fy,Lonzarich1986_G,Edwards1992_EL,Ishigaki1999_IM,Moriya2003_MU}. In view of our observation of low-energy excitations near $\mathbf{Q}_{\delta}$ indicating the proximity to magnetic order, here we test whether the large measured $\gamma$ is due to the spin fluctuations we observe. We can compare the measured magnetic response function  $\chi^{\prime\prime}(\mathbf{Q},\omega)$ to the SF-theory and hence estimate the heat capacity $\gamma$. This approach has been applied to a number of correlated electron systems \cite{Hayden2000hr}, most recently Sr$_3$Ru$_2$O$_7$ \cite{Lester2021_LRPC}. The SF-theory predicts that for $T \rightarrow 0$ (See Refs.~\cite{Lonzarich1986_G,Edwards1992_EL,Ishigaki1999_IM} and Methods), 
\begin{equation}
    \gamma_{\mathrm{SF}} = \frac{\pi k_B^2}{\hbar} \Big \langle\frac{1}{\Gamma(\mathbf{Q})}\Big \rangle_{\mathrm{BZ}},
    \label{eqn:gammma_SF}
\end{equation}
where $\Gamma(\mathbf{Q})$ is the spin relaxation rate and $\langle \ldots \rangle_{\text{BZ}}$ denotes average over the Brillouin zone.
We can use our fitted $\Gamma(\mathbf{Q})$ for the normal state at $T=T_c$ (plotted as $\Gamma^{-1}(\mathbf{Q})$ in Fig.\,\ref{fig:LET_E_cuts}f) to estimate $\gamma_{\mathrm{SF}}(T_c) = 16.5 \pm 3$ mJ mol$^{-1}$ K$^{-2}$ for LSCO ($p$ = 0.22), where we have corrected Eqn.\;\ref{eqn:gammma_SF} for finite temperatures (see Methods). The result agrees reasonably with the measured \cite{Matsuzaki2004_MMOI} value $\gamma_{\mathrm{exp}}(T_c) = 14.2$~mJ mol$^{-1}$ K$^{-2}$ (see Fig.~\ref{fig:entropy_heat_capacity}b). We also computed $\gamma_{\mathrm{SF}}$ for slightly underdoped LSCO ($p$ = 0.14) where the low-energy spin excitations at $T_c$ have previously been measured and parameterized  \cite{Aeppli1997jl}. Reasonable agreement between  $\gamma_{\mathrm{exp}}$ and $\gamma_{\mathrm{SF}}$ is also found (see Fig.\;\ref{fig:entropy_heat_capacity}b and Table\;\ref{tab:parms}).

%\textit{Superconducting State} 
In the superconducting state ($T$ = 2\;K, $B$ = 0 T), the low-energy excitations are suppressed leading to the gaped spectrum in Fig.~\ref{fig:LET_E_cuts}b (blue line) and there is a concomitant reduction of the specific heat \cite{Matsuzaki2004_MMOI,Girod2021_GLDS} with $\gamma_{\mathrm{exp}} \approx $ 4~mJ~mol$^{-1}$~K$^{-2}$.   Such a reduction is expected in the SF-theory on general grounds because of the suppression of the low-energy spin fluctuations. The application of magnetic field $B=8.8$\;T introduces low-energy excitations below the spin gap energy (Fig.~\ref{fig:LET_E_cuts}b, green line) which are associated with the vortices \cite{Lake2001_LACM}. In this inhomogeneous mixed state, the excitation spectrum is expected to be approximately a superposition of the contribution from the vortices which should be similar to the normal state (Fig.~\ref{fig:LET_E_cuts}a) and the $B=0$ spectrum in the superconducting state (Fig.~\ref{fig:LET_E_cuts}b, blue line). This is qualitatively consistent with the observed increase in specific heat with $\gamma_{\mathrm{exp}}$($T$ = 2\;K, $B$ = 8.8 T) $\approx$ 8.5~mJ~mol$^{-1}$ K$^{-2}$ \cite{Girod2021_GLDS}.  The normal-state-like component would continue to grow with increasing magnetic field leading to the large observed $\gamma_{\mathrm{exp}}$($T$ = 2\;K, $B$ = 34\;T) $\approx$ 15~mJ mol$^{-1}$ K$^{-2}$ near $B_{c2}$ \cite{Girod2021_GLDS}.    

Our results are complemented by recent x-ray diffraction measurements \cite{Miao2021_MFKM} of charge-density-wave (CDW) correlations in LSCO ($p=0.21$) with slightly lower doping.  These reveal CDW correlations with a propagation vector $\mathbf{Q}_{\mathrm{CDW}} = (\delta_{\mathrm{CDW}},0)$, where $\delta_{\mathrm{CDW}}\approx0.236$. As $\delta_{\mathrm{CDW}} \approx 2\delta_{\mathrm{SF}}$ it is likely that the CDW and spin fluctuations are related. However, NMR measurements \cite{Wu2015_WMKH} show that the CDW is frozen for frequencies $\hbar \omega \sim 0.5 \mu$eV$ \ll k_B T$ and the x-ray measurements show that the CDW has a much larger correlation length $\xi_{\mathrm{CDW}} =75 \pm 5$~\AA\ $\approx 4\xi_{\mathrm{SF}}$ \cite{Miao2021_MFKM}.  Thus it is unlikely that the CDW would give rise to a large specific heat $\gamma$ in a similar way to spin fluctuations.

\begin{table*}[ht]
\centering
\begin{tabular}{|l|l|l|l|l|l|l|l|l|l|} \hline
\multicolumn{1}{|c|}{Doping} & \multicolumn{1}{c|}{T} & \multicolumn{1}{c|}{B} & \multicolumn{1}{c|}{$\chi_{\delta}$} &  \multicolumn{1}{c|}{$\hbar\Gamma_{\delta}$} & \multicolumn{1}{c|}{$\xi^{-1}$} &   \multicolumn{1}{c|}{$\delta$} &
\multicolumn{1}{c}{$\gamma_{\textrm{exp}}$} &
\multicolumn{1}{c|}{$\gamma_{\textrm{SF}}$} &
\multicolumn{1}{c|}{$\langle m^2 \rangle$} \\
%\multicolumn{2}{c|}{$\gamma$} \\
 \multicolumn{1}{|c|}{p} & \multicolumn{1}{c|}{(K)} & \multicolumn{1}{c|}{(T)} & \multicolumn{1}{c|}{($\mu_B^2$ eV$^{-1}$ f.u.$^{-1}$)} & \multicolumn{1}{c|}{(meV)} & \multicolumn{1}{c|}{(\AA$^{-1}$)} & \multicolumn{1}{c|}{(r.l.u.)} & \multicolumn{2}{c|}{(mJ mole$^{-1}$\;K$^{-2}$) } &
 \multicolumn{1}{c|}{($\mu_B^2$ f.u.$^{-1}$)} \\  \hline
0.14  & 35 & 0 &  376(16) \cite{Aeppli1997jl} & 9.6(8) \cite{Aeppli1997jl} &  0.037(6) \cite{Aeppli1997jl} & 0.123 \cite{Aeppli1997jl} &  6.0 \cite{Mason1993_MAHR} & 5(1) & 0.18(2)\\
0.163 & 38.5 & 0   &   & 9 \cite{Lake2001_LACM} &         &           &   &  &\\   
0.22 & 26 & 0 & 71(15) & 4.6(3) & 0.053(8) & 0.134(4) & 14.2 \cite{Matsuzaki2004_MMOI} & 16.5(30) & 0.13(3) \\
0.22 & 2 & 0   &   &   &         &           & 4.0 \cite{Matsuzaki2004_MMOI,Girod2021_GLDS} & & \\     
0.22 & 2 & 8.8 &   &   & 0.057(9) & 0.135(4) & 8.5 \cite{Girod2021_GLDS} & & \\
0.22 & 2 & 34 &   &   &  &  & 15.0 \cite{Girod2021_GLDS} & & \\
\hline 
\end{tabular}
    \caption{\textbf{Parameterization the magnetic response and the heat capacity of LSCO.} Columns 4-7 show the parameterization of $\chi^{\prime\prime}(\mathbf{Q},\omega)$ using Eqn.~\ref{eqn:chi_MMP_SM} (normal state) and Eqn.~\ref{eqn:4} (superconducting state). Column 8 is the measured low-temperature linear heat capacity $\gamma=C/T$. Column 9 is $\gamma$ calculated from SF-theory using parameters in columns 5-7. Column 10 is the fluctuating moment squared in the range [-10,10] meV calculated using Eqn.~\ref{eqn:chi_MMP_SM}.} 
    \label{tab:parms}
\end{table*}

% location of QCP and pseudogap
We next consider how our data is related to the incipient incommensurate antiferromagnetism (ICAF) in the cuprates.  Previous measurements \cite{Aeppli1997jl} on LSCO ($p=0.14$) show an approximate $\omega/T$ scaling of the spin fluctuations consistent with a nearby ICAF-QCP.  Comparing these measurements for $T \approx T_c$ with our data, we find (Table~\ref{tab:parms}) that decreasing the doping from $p=0.22$ to 0.14, causes $\xi$ to increase from $19 \pm 2$ to $27 \pm 4$~\AA\ and $\chi_{\delta}$ to increase from $71 \pm 15$ to $376 \pm 16$~$\mu_B^2$ meV$^{-1}$ f.u.$^{-1}$.  We expect $\xi$ and $\chi_{\delta}$ to diverge approaching a QCP and the increases we observe suggest that the underlying QCP occurs for $p<0.22$.  We also expect the relaxation rate $\Gamma_{\delta}$ to decrease approaching a QCP as the fluctuations slow down. However, the opposite behaviour is observed, with $\hbar \Gamma_{\delta}$ increasing from $4.6 \pm 0.3$ to $9.6 \pm 0.8$~meV between $p=0.22$ and $p=0.14$. The obvious explanation for this is that the pseudogap which is present for $p<p^{\star}$ shifts the spectral weight to higher energy in the normal state \cite{Dai1999_DMHA} masking the critical behaviour in $\Gamma_{\delta}$. 

% behaviour in field
For $p=0.22$, we find that the low-energy $\chi^{\prime\prime}(\mathbf{Q}_{\delta},\omega)$ in the mixed state with $T \ll T_c$ and $B=8.8$~T is larger than in the normal state for $T \sim T_c$ indicating that $\chi^{\prime\prime}(\mathbf{Q}_{\delta},\omega)$ would continue to increase with decreasing $T$ if superconductivity did not intervene. Thus, there is a strong tendency to ICAF order in a magnetic field for $p=0.22$. If a larger field $B\sim B_{c2}$ were used to suppress superconductivity, we would expect that the low-energy spin response would be further enhanced (see Fig.~\ref{fig:LET_E_cuts}c,d).  The behaviour differs at lower doping, $p<p^{\star}$, where the application of field $B=7.5$~T at $T=7.7$~K$\ll T_c$ for $p=0.163$ does not lead to an enhancement of the low-energy $\chi^{\prime\prime}(\mathbf{Q}_{\delta},\omega)$ relative to the normal state at $T \sim T_c$ \cite{Lake2001_LACM}, presumably because of the presence of the pseudogap in field.  

% SF in overdoped cuprates
The low-energy collective spin fluctuations we observe for $p \! \gtrsim \! p^{\star}$ can account for the large low-temperature electronic heat capacity $\gamma$ in overdoped cuprates. La$_{2-x}$Sr$_x$CuO$_4$ \cite{Loram2001xp,Matsuzaki2004_MMOI, Momono1994_MINO,Girod2021_GLDS} and the related system La$_{2-x-y}$(Eu,Nd)$_y$Sr$_x$CuO$_4$ \cite{Michon2019fk} show a peak in $\gamma(p)$ close to $p^{\star}$ where the pseudogap disappears. For dopings $p<p^{\star}$, the pseudogap removes the low-energy electron quasiparticle states near $E_F$ and their spin degrees of freedom \cite{Timusk1999aa}. When these are restored near $p^{\star}$ they will contribute to the low-energy spin fluctuations, bring entropy and lead to the large $\gamma$.   The overdoped cuprate Tl$_2$Ba$_2$CuO$_{6+\delta}$, which shows no pseudogap, also shows large quasiparticle mass enhancements of $\sim 3$ for dopings $p \ge 0.27$ \cite{Rourke2010_RBBM} suggesting that the co-existence of low-energy spin fluctuations and large quasiparticle mass persists at large $p$ in other cuprate systems.

% T-linear
The existence of spin fluctuations with an energy scale comparable to temperature $\hbar \Gamma_{\delta} \approx k_B T$, such as those observed here, may be related to the strange metal or $T$-linear resistivity behaviour observed in overdoped cuprates. This has been described in terms of ``Planckian dissipation'' \cite{Keimer2015_KKNU,Cooper2009aa,Hartnoll2021_HM} where the the inverse  Planckian time varies as $\tau^{-1}_{\textrm{Pl}} \approx k_B T/\hbar$.  The low-energy spin fluctuations in underdoped cuprates are strongly $T$-dependent, with $\Gamma_{\delta}$ increasing rapidly with temperature up to 300~K \cite{Aeppli1997jl}. Our work suggests that a similar strong  $T$-dependence may also persist in the overdoped region for $T>T_c$. However, further work is required to establish this.

\section{Methods}
\textbf{Single crystal sample growth and characterisation.}
Single crystals of La$_{2-x}$Sr$_x$CuO$_4$ ($x$ = 0.22) were grown by the travelling-solvent floating-zone method. 
The crystals were annealed in 1 bar of flowing oxygen at 800 $^{\circ}$C for 6 weeks. The Sr concentration was determined by SEM-EDX and ICP-AES to be $x$ = 0.215 $\pm$ 0.005. SQUID magnetometry measurements show that $T_{c,\text{onset}}$ = 26 K. Previous INS measurements on these crystals \cite{Lipscombe2007_LHVM} have shown a double-peak structure ($\sim$10 and 120 meV) in the local susceptibility $\chi^{\prime\prime}(\omega)$ in the superconducting state.

\textbf{Inelastic neutron scattering.}
Inelastic and elastic neutron scattering experiments were performed using the direct-geometry time-of-flight spectrometers LET and MERLIN at ISIS Facility and the IN12 triple-axis spectrometer at the Inititut Laue-Langevin. Co-aligned single crystals with a total masses of 29.8, 32.7 and 5.9 grams were used at LET, MERLIN and IN12, respectively.  
At LET, the $c$ axis was mounted vertically and the data were collected by rotating the samples in 1 degree steps using incident neutron energies $E_i$ = 3.3, 4.3, 10 and 16 meV, in the high-flux mode with the resolution chopper frequency set to 120 Hz, giving rise to elastic energy resolutions of $\sim$0.12, 0.18, 0.6 and 1.2 meV, respectively. A vertical magnetic field up to 8.8 T was applied along the $c$ axis. At MERLIN, we used E$_i$ = 30 meV and the chopper frequency 150 Hz, giving rise to an elastic energy resolution of $\sim$1.8 meV, and the samples were oriented with [110] and [001] directions in the horizontal scattering plane. At IN12, a vertical magnetic field up to 10 T was applied along the $c$ axis. The data were collected using a fixed final energy $E_f$ = 4.7 meV, collimation of open-80$^{\prime}$-open-open, a horizontally focused pyrolytic graphite analyser  and a velocity selector in the incident beam. Our initial observation of the low-energy spin fluctuations and the field-induced response was made on IN12 (See Extended Data Fig.~\ref{fig:IN12_data}). 

LSCO ($x$ = 0.22) has a tetragonal structure with $a = b \approx 3.77$ \AA~and $c \approx 13.18$ \AA.  We label the reciprocal space as $\mathbf{Q} = H \mathbf{a}^{\star} + K \mathbf{b}^{\star} + L\mathbf{c}^{\star}  \equiv (H,K,L)$.  The scattered intensity has been scaled to absolute units using the incoherent scattering of vanadium.  Comparison of absolute intensities between different samples and instruments requires sample absorption and illumination to be taken into account. We estimate the error in our normalisation is about $\pm 20$\%. This is reflected in the error of $\chi_{\delta}$.   
The scattering cross-section is related to the scattering function $S(\mathbf{Q},\omega)$ and energy- and wavevector-dependent magnetic response function $\chi^{\prime\prime}(\mathbf{Q},\omega)$ by the fluctuation-dissipation theorem
\begin{eqnarray}
 \frac{k_i}{k_f}\frac{d^2\sigma}{d\Omega\, dE} &=& S(\mathbf{Q},\omega) \\ & = & \frac{2(\gamma r_e)^2}{\pi g^2 \mu_B^2}\vert F(\mathbf{Q})\vert^2\frac{\chi^{\prime\prime}(\mathbf{Q},\omega)}{1-\textrm{exp}(-\hbar\omega/k_B T)},
\label{eqn:1}
\end{eqnarray}
where $(\gamma r_e)^2$ = 0.2905 barn sr$^{-1}$, and $F(\mathbf{Q})$ the magnetic form factor. 
The data were fitted using Eqn.~\ref{eqn:4} and convoluted with the instrumental resolution using the Horace package \cite{Ewings2016_EBLD}.  

\textbf{Data fitting.} In order to obtain a global fit of Eqn.~\ref{eqn:chi_MMP_SM} to a set of constant-energy cuts through the normal-state data at $T=T_c$ such as those in Extended Data Fig.~\ref{fig:fit_parms}, we fit the individual cuts to the form 
\begin{equation}
    \chi^{\prime\prime}(\mathbf{Q},\omega) = \frac{\chi^{\prime\prime}(\mathbf{Q}_{\delta},\omega)[\xi^{-4}+\kappa_1^4(\omega)]}{[\xi^{-2}+R(\mathbf{Q})]^2+\kappa_1^4(\omega)},
\label{eqn:4}
\end{equation}
where $\chi^{\prime\prime}(\mathbf{Q_{\delta}},\omega)$ and $\kappa_1^2(\omega)$ vary as 
\begin{equation}
    \chi^{\prime\prime}(\mathbf{Q}_{\delta},\omega) = \frac{\chi_{\delta}(\omega/\Gamma_{\delta})}{ 1+(\omega^2/\Gamma_{\delta}^2)}
    \label{eqn:5}
\end{equation}
and
\begin{equation}
    \kappa_1^2(\omega)=\frac{\omega}{\Gamma_{\delta}}\xi^{-2}
    \label{eqn:6}
\end{equation}
to reproduce Eqn.~\ref{eqn:chi_MMP_SM} \cite{Hayden2000hr}. 
We obtain $\xi^{-1}=0.053 \pm 0.008$~\AA$^{-1}$ from fitting the $\hbar\omega$ = 1.25 meV data. The solid lines in Extended Data Fig.~\ref{fig:fit_parms}a-c show fits of Eqn.~\ref{eqn:4} to the constant-energy cuts of $S(\mathbf{Q},\omega)$.  Extended Data Fig.~\ref{fig:fit_parms}d-f show the fitting parameters $\chi^{\prime\prime}(\mathbf{Q}_{\delta},\omega)$, $\kappa_1^2(\omega)$, and $\delta$ as a function of energy. 
The fitted values of $\chi^{\prime\prime}(\mathbf{Q}_{\delta},\omega)$ are well described by Eqn.~\ref{eqn:5}, with $\hbar \Gamma_{\delta}$ = 4.3 $\pm$ 0.3 meV and $\chi_{\delta}$ = 71 $\pm$ 15 $\mu_B^2 \text{eV}^{-1} \text{f.u.}^{-1}$. The value of $\Gamma_{\delta}$ agrees with that obtained directly from the raw data (Fig.~\ref{fig:LET_E_cuts}a). The gradient of Fig.~\ref{fig:fit_parms}e yields 6.4$\times 10^{-4}$~ \AA$^{-2}$~meV$^{-1}$ close to the expected value $\xi^{-2}/(\hbar\Gamma_{\delta})=6.1\times 10^{-4}$~\AA$^{-2}$~meV$^{-1}$ according to Eqn.~\ref{eqn:6}. 
Thus, the fitting procedure is self-consistent demonstrating that the magnetic response in the normal state of LSCO ($p$ = 0.22) at $T_c$ is well described by Eqns.~\ref{eqn:chi_MMP_SM} and \ref{eqn:R_Q}.

In the superconducting state, the line shape of the energy-dependent magnetic response at $\mathbf{Q}=\mathbf{Q}_{\delta}$ (Fig.~\ref{fig:LET_E_cuts}b) is no longer described by a Lorentzian form. The constant-energy cuts (Fig.~\ref{fig:LET_q_cuts}d-f) were fitted by a model used by Aeppli \textit{et al.} in Ref.~\cite{Aeppli1997jl} which is equivalent to Eqn.~\ref{eqn:4} with $\kappa_1(\omega)=0$.

\textbf{Spin fluctuation heat capacity model.}
The heat capacity of nearly antiferromagnetic metals can be understood within the self-consistent renormalisation spin fluctuation theory (SF-theory) based on a one-loop approximation and the Hubbard model \cite{Beal-Monod1968na,Brinkman1968fy,Lonzarich1986_G,Edwards1992_EL,Ishigaki1999_IM,Moriya2003_MU}. The low-temperature free energy $F$ can be expressed as 
\begin{align}
    F = \sum_{\mathbf{Q}} \int_{0}^{\omega_{\text{c}}} \; d\omega  F_{\text{osc}}(\omega,T) \frac{3}{\pi} \frac{\Gamma(\mathbf{Q})}{\omega^2+\Gamma^2(\mathbf{Q})}, \label{eqn:free_energy}
\end{align}
where $F_{\text{osc}}(\omega,T)=\hbar \omega/2+ k_B T \ln[1-\exp(-\hbar \omega/ k_B T)]$ is the free energy of a harmonic oscillator with frequency $\omega$ and $\Gamma(\mathbf{Q})$ is the relaxation rate of a spin fluctuation mode of wavevector $\mathbf{Q}$.  The $\Gamma(\mathbf{Q})/[\omega^2+\Gamma^2(\mathbf{Q})]$ factor is proportional to  $\chi^{\prime\prime}(\mathbf{Q},\omega)/\omega$. When combined with the zero-point energy $\hbar \omega/2$ in $F_{\text{osc}}$, it yields a $1/\omega$ high-energy tail in the integrand and logarithmic singularity in the integral. The tail is unphysical, it arises because of the use a Lorentzian description of $\chi^{\prime\prime}(\mathbf{Q},\omega)/\omega$. In reality, the tail would drop off faster than $1/\omega$. To avoid introducing unphysical states into $F$ as $\omega \rightarrow \infty$, a cut-off frequency $\omega_{\text{c}}$ is introduced \cite{Ishigaki1999_IM} where $\omega_{\text{c}} \gg \Gamma$ (See Supplementary Material for discussion). 

Eqn.~\ref{eqn:free_energy} may be used to obtain an approximate expression \cite{Lonzarich1986_G,Edwards1992_EL,Ishigaki1999_IM} for the linear coefficient of the specific heat $\gamma$,
\begin{align}
\gamma_{\text{SF}} &= \frac{C}{T} = -\frac{\partial^2 F}{\partial T^2}  \\
&= \sum_{\mathbf{Q}} \; \int_{0}^{\omega_{\mathrm{c}}} \; d\omega \; \frac{C_{\text{osc}}(\omega,T)}{T} \frac{3}{\pi} \frac{\Gamma(\mathbf{Q})}{\Gamma(\mathbf{Q})^2+\omega^2}. \label{eqn:gamma}
\end{align}
This is the sum of the specific heat $C_{\text{osc}}$ of oscillators with a frequency distribution of oscillator frequency $\omega$ corresponding to the response function $\chi^{\prime\prime}(\mathbf{Q},\omega)/\omega$, where
\begin{align}
C_{\text{osc}}(\omega,T) = \frac{\hbar^2 \omega^2}{k_B T^2}  \frac{e^{\hbar\omega/k_B T}}{(e^{\hbar \omega/k_B T}-1)^2}. \label{eqn:C_osc}
\end{align}
The low-temperature limit of Eqn.~\ref{eqn:gamma} is Eqn~\ref{eqn:gammma_SF}. 
Eqn.~\ref{eqn:gamma} does not have the $1/\omega$ tail described above, instead the integrand $\sim e^{-\omega}$ for $\omega \rightarrow \infty$. The function $C_{\text{osc}}(\omega,T)$ acts to pick out the excitations $\lesssim 2 k_B T$ and as a high-frequency cut off.   If $\Gamma(\mathbf{Q})$ is known, Eqn.~\ref{eqn:gamma} can be evaluated numerically by taking the limit $\omega_{\text{c}} \rightarrow \infty$ (See Table~\ref{tab:parms}).  Our calculated $\gamma_{\text{SF}}$ is predominately due to spin fluctuations in the energy range [0,10] meV probed by our experiment and not strongly dependent on $\omega_{\text{c}}$ for $\omega_{\text{c}} >$10~meV. We find that  
$\gamma_{\text{SF}}(\omega_{\text{c}}=10~\text{meV})/\gamma_{\text{SF}}(\omega_{\text{c}} \rightarrow \infty) =0.93$ for $T=26$~K (See Supplementary Material). 
\section{Data availability}

Data collected at ISIS on LET and MERLIN are available at \url{https://doi.org/10.5286/ISIS.E.RB1920542} and
\url{https://doi.org/10.5286/ISIS.E.RB2010576}.
Data collected at the ILL on IN12 are available at \url{https://doi.ill.fr/10.5291/ILL-DATA.4-02-583}.

\section{Computer Code}
Mathematica computer code used to evaluate $\gamma_{\textrm{SF}}$ in Table~\ref{tab:parms} is available with the paper. 

\section{acknowledgments}
We are grateful to J. R. Stewart for running the LET experiment. We acknowledge useful discussions with A. Carrington and N. E. Hussey. M.Z. and S.M.H. acknowledge funding and support from the Engineering and Physical Sciences Research Council (EPSRC) under Grant No.~EP/R011141/1. 
%We acknowledge ISIS Facility for instrument time at beamline LET under proposal 1920542. %Merlin under proposal 2010576 and Institut Laue-Langevin for time at IN12 with proposal No. 4-02-561.

\section{author contributions}
M.Z. and O.J.L. prepared the samples. M.Z., D.J.V., S.R., C.C.T. and S.M.H. made neutron scattering measurements.  M.Z. and S.M.H. analyzed the data and wrote the initial manuscript. All authors contributed to the discussion and provided feedback on the manuscript.

\section{competing financial interests}
The authors declare no competing financial interests.

\renewcommand{\tablename}{Extended Data Table}

\renewcommand{\figurename}{Extended Data Figure} 

\setcounter{figure}{0}

\begin{figure*}[h]
\includegraphics[width = \textwidth]{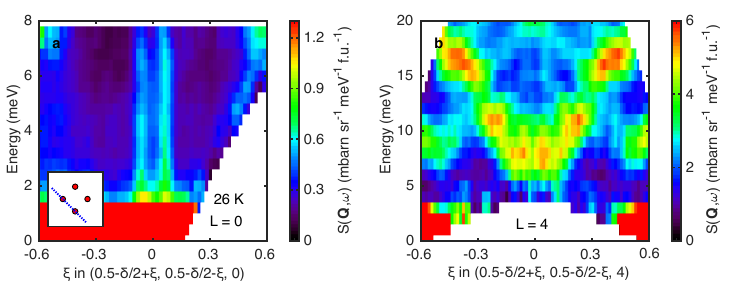}
\caption{\textbf{Spin fluctuations and phonons in  La$_{2-x}$Sr$_x$CuO$_4$ ($x = 0.22$) near $\mathbf{Q_{\delta}}$.}
$S(\mathbf{Q},\omega)$ as a function of energy and wavevector along a trajectory through two incommensurate wave vectors $\mathbf{Q_{\delta}}$ = (0.5-$\delta$, 0.5, $L$) and (0.5, 0.5-$\delta$, $L$) (see inset to panel \textbf{a}). Integration ranges are  \textbf{a} $L \in [-1,1]$ and \textbf{b} $L \in [3.8,4.2]$. Strong phonons are observed (panel \textbf{b}) for $L \approx$  4, but these are not visible near $L=0$ (panel \textbf{a}) where spin fluctuations are seen. Data were collected on LET (panel \textbf{a}) and MERLIN (panel \textbf{b}).
\label{fig:LET_MERLIN}}
\end{figure*}

\begin{figure*}[h]
\includegraphics[width = \textwidth]{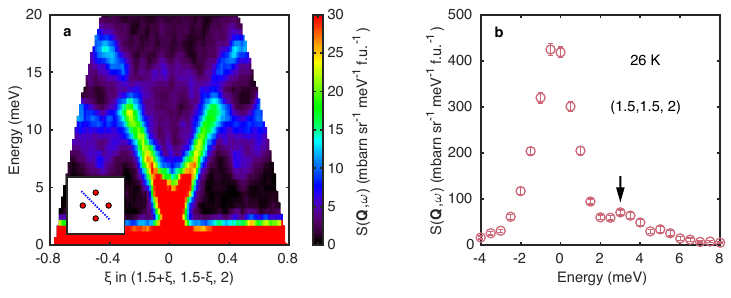}
\caption{\textbf{Phonons in  La$_{2-x}$Sr$_x$CuO$_4$ ($x = 0.22$) near (1.5, 1.5, 2).}
\textbf{a} $S(\mathbf{Q},\omega)$ as a function of energy and wavevector across (1.5, 1.5, 2) with $L \in [1.8,2.2]$ at $T$ = 26 K. \textbf{b} Energy dependence of $S(\mathbf{Q},\omega)$ at (1.5, 1.5, 2). The arrow denotes a phonon at $\sim$3 meV corresponding to the rotation of the CuO$_6$ octahedra. These features are quite different from the scattering we observe near (0.5, 0.5, 0) identified as magnetic scattering. Data were collected on MERLIN.
\label{fig:MERLIN_phonons}}
\end{figure*}

\begin{figure*}[h]
\includegraphics[width = 0.8\textwidth]{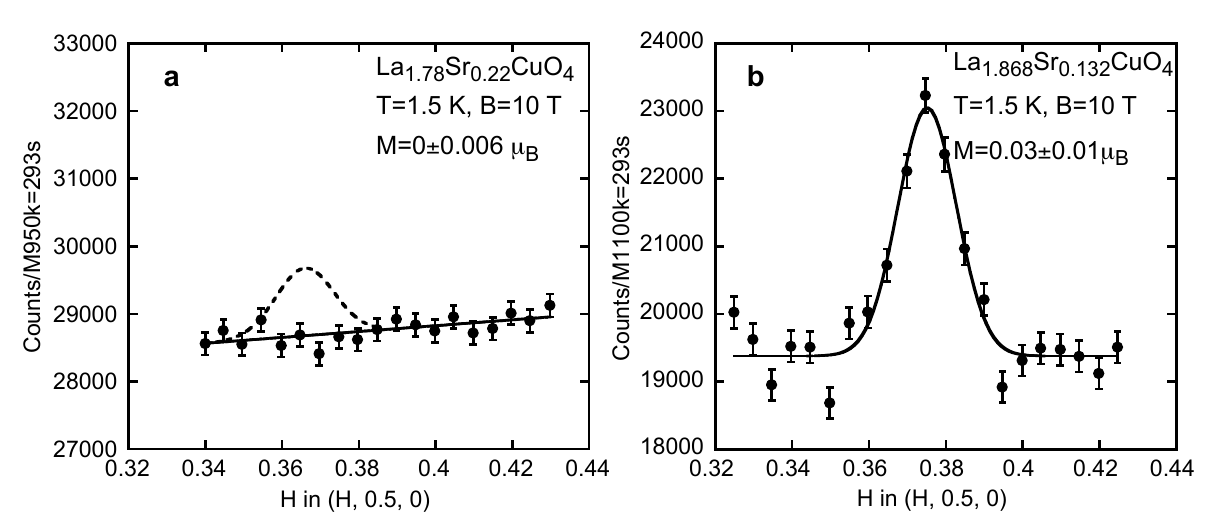}
\caption{\textbf{No evidence for field-induced spin density wave (SDW) order in LSCO ($p=0.22$)}.   Elastic scans through the $\mathbf{Q}_{\delta}=(0.5-\delta, 0.5, 0)$ position collected on IN12 with $E_f=4.7$ meV, $T=1.5$~K and $B=10$~T. \textbf{a} No evidence of SDW order is seen in the La$_{2-x}$Sr$_x$CuO$_4$ ($x = 0.22$) sample studied here. The dashed line shows the position and width (due to instrumental resolution) that a SDW peak at $\mathbf{Q}_{\delta}$ would have. \textbf{b} For comparison, we show a SDW peak measured on an underdoped La$_{2-x}$Sr$_x$CuO$_4$ ($x = 0.132$) sample of similar size with IN12 using similar spectrometer conditions and scaled to the same time as  \textbf{a}. The values of ordered moments are for a single $\mathbf{Q}_{\delta}$ and have been determined by comparison with scattering from the (110) Bragg peak.}  
\label{fig:IN12_noSDW}
\end{figure*}

\begin{figure*}[h]
\includegraphics[width = \textwidth]{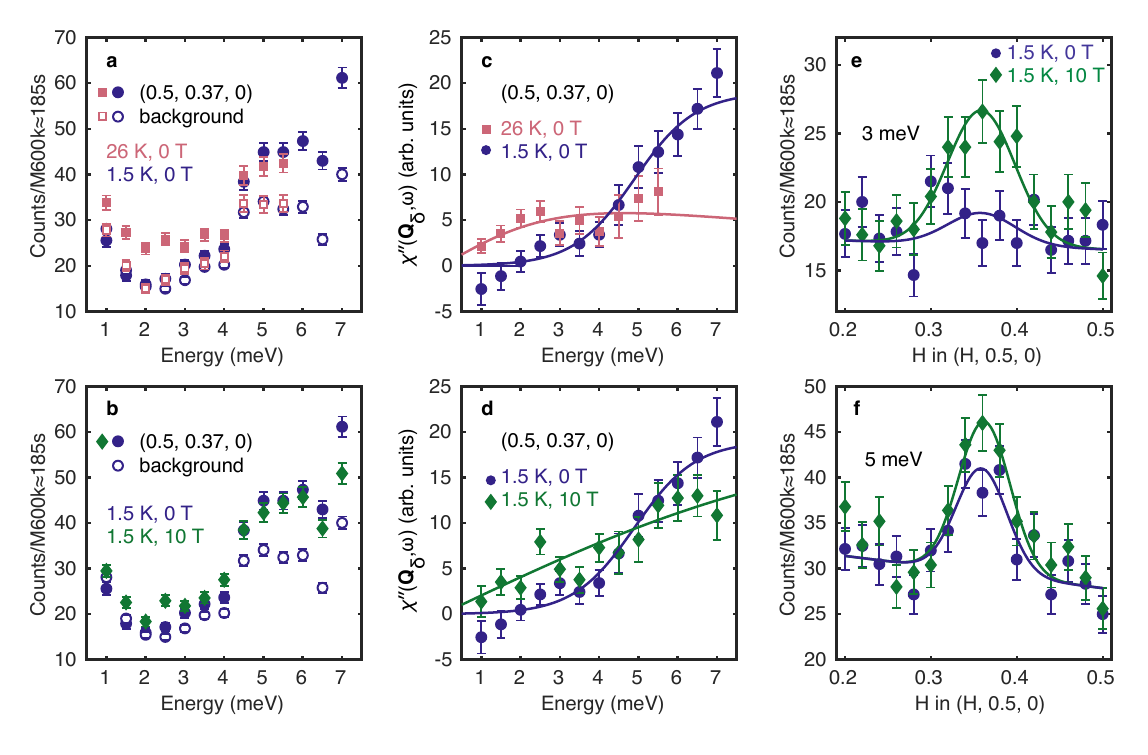}
\caption{\textbf{Low-energy spin fluctuations measured by IN12 cold neutron triple-axis spectrometer.}  \textbf{a-b} Measurements made at $\mathbf{Q}_{\delta}$ = (0.5, 0.37, 0) (closed symbols) and a background estimated from the average of (0.56, 0.31, 0) and (0.44, 0.43, 0) (open symbols). \textbf{c-d} Signal isolated from the data in \textbf{a}-\textbf{b} and corrected by a bose factor.  Data are consistent with the LET data and show low-energy spin fluctuations in the normal state (panel \textbf{c}) and a field-induced signal in the superconducting state (panels \textbf{d}). \textbf{e-f} Constant-energy scans across $\mathbf{Q}_{\delta}$ at $T$ = 1.5 K, $B$ = 0 and 10 T.    
\label{fig:IN12_data}}
\end{figure*}

\begin{figure*}[h]
\includegraphics[width = \textwidth]{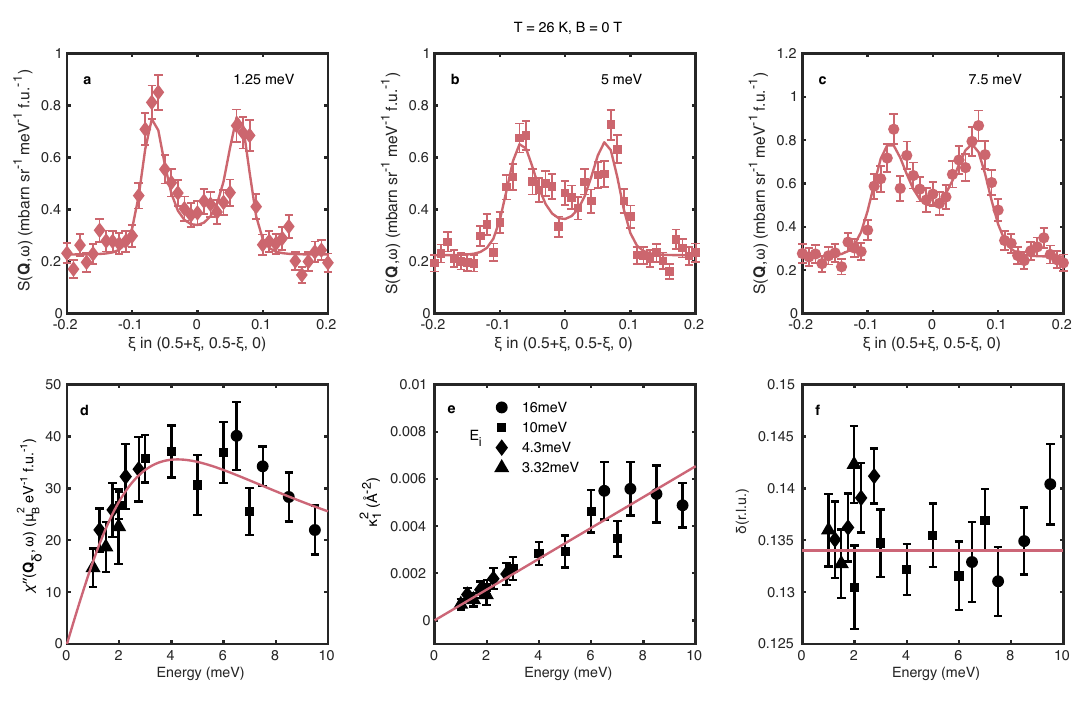}
\caption{\textbf{Fits of low-energy spin fluctuations in the normal state at $T_c$.} \textbf{a-c} Constant-energy cuts of $S(\mathbf{Q},\omega)$. Integration range perpendicular to the trajectory is shown in Fig.~\ref{fig:LET_slices}h by dashed lines and also $|L| \le 1$. Solid lines are fitted curves using Eqn.~\ref{eqn:4} convoluted with the instrumental resolution.
\textbf{d-f} Energy dependence of $\chi^{\prime\prime}(\mathbf{Q}_{\delta},\omega)$, $\kappa_1^2(\omega)$, and $\delta$ in Eqn.~\ref{eqn:4}. The solid lines in \textbf{d}, \textbf{e} and \textbf{f} are fits of Eqn.~\ref{eqn:5}, Eqn.~\ref{eqn:6} and a constant.  
\label{fig:fit_parms}}
\end{figure*}

\bibliography{LSCO22_LET_min2}

\end{document}

\title{Spin fluctuations associated with the collapse of the pseudogap in a cuprate superconductor: Supplementary Material}

\author{M. Zhu}
\affiliation{H.H. Wills Physics Laboratory, University of Bristol, Tyndall Avenue, Bristol BS8 1TL, United Kingdom}

\author{D. J. Voneshen}
\affiliation{ISIS Facility, Rutherford Appleton Laboratory, Didcot OX11 0QX, United Kingdom}
\affiliation{Department of Physics, Royal Holloway University of London, Egham, TW20 0EX, United Kingdom}
\author{S. Raymond}

\affiliation{Univ. Grenoble Alpes, CEA, IRIG, MEM, MDN, 38000 Grenoble, France}
\author{O. J. Lipscombe}
\affiliation{H.H. Wills Physics Laboratory, University of Bristol, Tyndall Avenue, Bristol BS8 1TL, United Kingdom}
\author{C. C. Tam}
\affiliation{H.H. Wills Physics Laboratory, University of Bristol, Tyndall Avenue, Bristol BS8 1TL, United Kingdom}
 \affiliation{Diamond Light Source, Harwell Campus, Didcot OX11 0DE, United Kingdom.}
\author{S. M. Hayden}
\affiliation{H.H. Wills Physics Laboratory, University of Bristol, Tyndall Avenue, Bristol BS8 1TL, United Kingdom}

\maketitle

% \section{Sample growth \& characterisation}

\section{High-frequency cut-off in spin fluctuation theory}

The spin-fluctuation theory of Refs.\cite{Beal-Monod1968na,Brinkman1968fy,Lonzarich1986_G,Edwards1992_EL,Ishigaki1999_IM,Moriya2003_MU} discussed in the main text requires a high-frequency cut-off $\omega_c$ when certain physical quantities are calculated. In the following we review this in order to place the present work in context. The aim of our work is to model the low-energy magnetic excitations and connect them to the thermal properties of the system. 

The damped harmonic oscillator (DHO) is an excellent starting point for understanding the response functions of physical systems.  We use the mechanical case for illustration and following the notation of Ref.~\cite{Chaikin1995} (Note $\gamma$ is different from the variable used to represent the linear heat capacity elsewhere in this paper). In the presence of a viscous force and external force $f$, the DHO equation is:
\begin{equation}
    \ddot{x}+\omega_0^2 x + \gamma \dot{x} = f/m, \label{Eq:DHO}
\end{equation}
where $\omega_0$ is the undamped oscillation frequency, $\gamma$ is a characteristic (viscous) inverse decay time, $x$ the particle displacement and $m$ its mass. We define the frequency-dependent response to a force as $\chi(\omega)=x(\omega)/f(\omega)$. It can be shown that 
\begin{equation}
    \chi^{\prime\prime}(\omega) \propto \frac{\omega \gamma}{(\omega^2-\omega^2_0)^2+(\omega \gamma)^2}.
    \label{Eq:DHO_chi}
\end{equation}
In the ``extreme overdamped limit'', $\gamma/2 \gg \omega_0$, the response of the DHO has two decaying components with inverse decay times
\begin{align}
    \tau_{f}^{-1} & \rightarrow \gamma, \\
    \tau_{s}^{-1} & \rightarrow \omega^2_0/\gamma.    
\end{align}
This is discussed in detail in Ref.~\cite{Chaikin1995} and other places. The fast decay time $\tau_f$ is much shorter than the slow decay time $\tau_s$. Thus for times long compared to $\tau_f$ (i.e. the low frequency limit of interest in the present experiment), the fast mode can be neglected. This approximation corresponds to neglecting the inertial term $m \ddot{x}$ in Eqn.~\ref{Eq:DHO}. The resulting equation of motion is
\begin{equation}
\gamma \dot{x} + \omega_0^2 x  = f/m, \label{Eq:EOHO}
\end{equation}
with the response function in the so-called ``extreme overdamped limit'' \cite{Chaikin1995} given as
\begin{equation}
\frac{\chi^{\prime\prime}(\omega)}{\omega} \propto \frac{\Gamma}{\omega^2+\Gamma^2}, 
\label{Eq:EOHO_chi}
\end{equation}
where the right hand side is a Lorentzian function and
\begin{equation}
\Gamma=\tau_s^{-1}=\omega_0^2/\gamma.  \label{Eq:Gamma_tau}
\end{equation}

In our work, the low-frequency susceptibility of a given $\mathbf{Q}$ mode is modelled with the extreme overdamped oscillator response function above \cite{Chaikin1995}
\begin{equation}
\chi^{\prime\prime}(\mathbf{Q},\omega)= \frac{\chi(\mathbf{Q}) \Gamma(\mathbf{Q}) \omega}{\Gamma^2(\mathbf{Q})+\omega^2}.
\label{eqn:chi_lorentzian}
\end{equation}
The fluctuation-dissipation theorem implies that the corresponding fluctuating moment squared due to this mode is given by \cite{Hayden2008book}
\begin{align}
\langle m^2(\mathbf{Q})\rangle = &
\lim_{\omega_c \rightarrow \infty} \int_{-\omega_{\text{c}}}^{\omega_{\text{c}}}  \frac{3}{\pi} \chi^{\prime\prime}(\mathbf{Q},\omega) \times \frac{1}{1-e^{-\hbar\omega/k_B T}} \,  d\omega \\
=& \lim_{\omega_c \rightarrow \infty} \int_{-\omega_{\text{c}}}^{\omega_c} \frac{3}{\pi} \frac{\chi(\mathbf{Q}) \Gamma(\mathbf{Q}) \omega}{\Gamma^2(\mathbf{Q})+\omega^2}  \times \frac{1}{1-e^{-\hbar\omega/k_B T}} \, d\omega. \label{eqn:m_sqr_2}
\end{align}
The use of the the Lorentzian response (Eqn.~\ref{eqn:chi_lorentzian}) means that the integral in Eqn.~\ref{eqn:m_sqr_2} will have a logarithmic divergence $\propto  \! \log \omega_{\text{c}}$ as $\omega_{\text{c}} \! \rightarrow \!  \infty$. This unphysical result arises because of the use of the Lorentzian form which has a high-frequency dependence $\chi^{\prime\prime}(\omega) \propto 1/\omega$. In reality, the high-energy response drops off more rapidly, for example in the DHO case above, it varies as $1/\omega^3$ at the highest frequencies where the $\omega^2$ term dominates the denominator of Eqn.~\ref{Eq:DHO_chi} i.e. $\omega \gtrsim \gamma$.   
 
Our approach is designed to understand the low-frequency ($\omega \sim \Gamma$) excitations whose population changes rapidly with temperature.  As shown in the paper, the low-frequency part of the response is well described by the Lorentzian without including additional fitting parameters. In order to exclude the unphysical high-energy states of the Lorentzian a high-frequency cut off  $\omega_\text{c}$ (where $\omega_{\text{c}} \sim \gamma$) can be introduced (see e.g. Ref.~\cite{Ishigaki1999_IM} and references therein).     

Our heat capacity is based on the free energy 
\begin{align}
    F = \sum_{\mathbf{Q}} \int_{0}^{\omega_{\text{c}}} \; d\omega  F_{\text{osc}}(\omega,T) \frac{3}{\pi} \frac{\Gamma(\mathbf{Q})}{\omega^2+\Gamma^2(\mathbf{Q})}, \label{eqn:free_energy}
\end{align}
where $F_{\text{osc}}(\omega,T)=\hbar \omega/2+ k_B T \ln[1-\exp(-\hbar \omega/ k_B T)]$ and the factor $\Gamma(\mathbf{Q})/[\omega^2+\Gamma^2(\mathbf{Q})]$ is proportional to  $\chi^{\prime\prime}(\mathbf{Q},\omega)/\omega$. In this case the zero-point fluctuations combined with the unphysical high-energy tail of the Lorentzian also leads to a $\log \omega_{\text{c}}$ term which can be suppressed with a high-frequency cut off.  

 Ishigaki and Moriya \cite{Ishigaki1999_IM} and the references therein show that the linear heat capacity obtained from Eqn.~\ref{eqn:free_energy} is
\begin{equation}
\gamma_{\text{SF}} = \sum_{\mathbf{Q}} \; \int_{0}^{\omega_{\mathrm{c}}} \; d\omega \; \frac{C_{\text{osc}}(\omega,T)}{T} \frac{3}{\pi} \frac{\Gamma(\mathbf{Q})}{\Gamma(\mathbf{Q})^2+\omega^2}. \label{eqn:gamma}
\end{equation}
This is the sum of the specific heat $C_{\text{osc}}$ of oscillators with a frequency distribution of oscillator frequency $\omega$ corresponding to the response function $\chi^{\prime\prime}(\mathbf{Q},\omega)/\omega$, where
\begin{equation}
C_{\text{osc}}(\omega,T) = \frac{\hbar^2 \omega^2}{k_B T^2}  \frac{e^{\hbar\omega/k_B T}}{(e^{\hbar \omega/k_B T}-1)^2}. \label{eqn:C_osc}
\end{equation}
The integrand of Eqn.~\ref{eqn:gamma} $\sim e^{-\omega}$ as $\omega \rightarrow \infty$ this means Eqn.~\ref{eqn:gamma} converges to a finite value as $\omega_{\text{c}} \rightarrow \infty$. In Supplementary Fig.~\ref{fig:LSCO_Heat_Capacity_Calc_with_cut_off} we show $\gamma_{\text{SF}}$ as a function of $\omega_{\text{c}}$ evaluated using Eqn.~\ref{eqn:gamma} with our fitted model parameters $\{\delta, \xi^{-1}, \Gamma_{\delta}\}$ for $T=26$~K (see Table~1). For $\hbar \omega_{\text{c}}=10$~meV, $\gamma^{\text{SF}}$ is 93\% of the value obtained in the $\omega_{\text{c}} \rightarrow \infty$ limit. This demonstrates that, within the model presented in this paper, the spin fluctuations responsible for the large specific heat in the normal state at $T=26$~K are predominately those measured in the energy range $0 \leq \hbar \omega \leq 10$~meV, not contributions from the high-energy tail. In Supplementary Fig.~\ref{fig:C_osc_fig}, panel (b) shows how $C_{\text{osc}}(\omega)$ suppresses the high-energy tail of the Lorentzian in panel (a) leading to a convergent integral where the heat capacity $\gamma$ is predominately due to excitations with energies $\lesssim 2 k_B T$.
\begin{center}
\includegraphics[width=0.7 \textwidth]{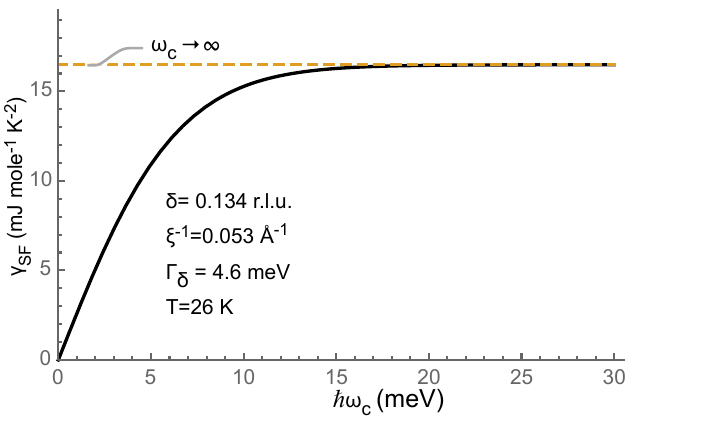}
\captionof{figure}{Calculation of the contribution of spin fluctuations to the linear heat capacity $\gamma_{\text{SF}}$ as a function of cut-off energy $\hbar \omega_{\text{c}}$ evaluated using Eqn.~\ref{eqn:gamma}.}
\label{fig:LSCO_Heat_Capacity_Calc_with_cut_off} 
\end{center}
\begin{center}
\includegraphics[width=0.9 \textwidth]{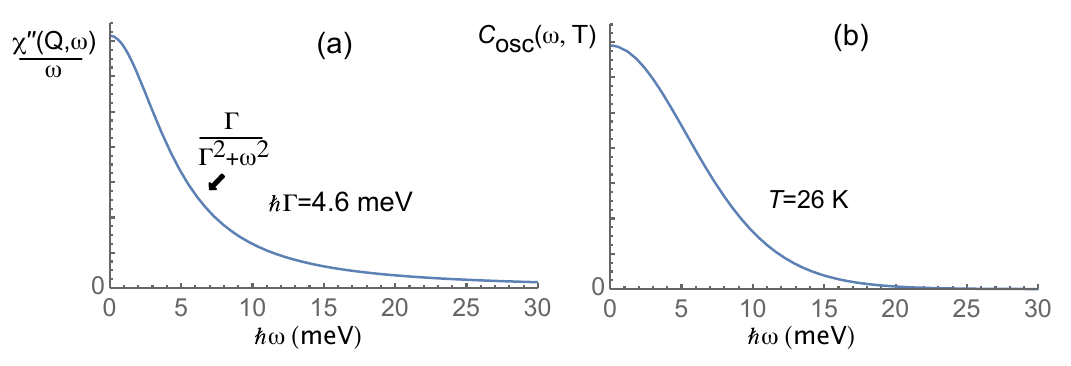}
\captionof{figure}{Plot of the two factors in the integrand of Eqn.~\ref{eqn:gamma}. Panel (b) shows how $C_{\text{osc}}(\omega)$ suppresses the high-energy tail of the Lorentzian in panel (a) leading to a convergent integral where the heat capacity $\gamma$ is predominately due to excitations with energies $\lesssim 2 k_B T$.}
\label{fig:C_osc_fig} 
\end{center}\bibliography{LSCO22_LET_min2}